\documentclass{jetpl}
\twocolumn

\rus

\usepackage{caption}
\usepackage{graphicx}
\usepackage{subcaption}
\usepackage{cite}
\usepackage[dvips]{graphicx}
\usepackage{multirow}
\DeclareGraphicsExtensions{.pdf,.png,.jpg,.eps,.ps}
\RequirePackage{caption}
\DeclareCaptionLabelSeparator{point}{. }
\makeatletter
\newcommand{\keepwithnext}{\@beginparpenalty 10000}
\makeatother

\title{Необычные события в экспериментах с рентген-эмульсионными камерами\\
The unusual events in X-ray emulsion chamber experiments}
\rtitle{Необычные события в экспериментах с рентген-эмульсионными камерами}
\sodtitle{Необычные события в экспериментах с рентген-эмульсионными камерами}

\author{С.\,Е.\,Пятовский\\S.\,E.\,Pyatovsky \/\thanks{e-mail: vgsep@ya.ru}}
\rauthor{С.\,Е.\,Пятовский}
\sodauthor{Пятовский}

\address{Физический институт им.~П.~Н.~Лебедева Российской академии наук, Москва, Россия\\
Lebedev Physical Institute, Russian Academy of Sciences, Moscow, Russia}

\abstract{Рассмотрены события, зарегистрированные в эксперименте с рентген-эмульсионными камерами (РЭК) по изучению стволов широких атмосферных ливней (ШАЛ) с энергиями ядер первичного космического излучения (ПКИ) $E_0$~>~100~ТэВ и не имеющие к настоящему времени своего объяснения в рамках стандартной модели ядерных взаимодействий (СМ). Метод РЭК единственный, позволяющий изучать стволы ШАЛ с высоким координатным разрешением $\sim$~100~мкм. Стволы ШАЛ содержат максимально полную информацию о первичных актах ядерных взаимодействий, вследствие чего все необычные события, регистрируемые методом РЭК, представляют большой научный интерес. Несколько типов необычных событий, зарегистрированных методом РЭК, удалось объяснить в рамках СМ.\newline
\textbf{Ключевые слова:} космические лучи, широкие атмосферные ливни, экзотические события в ядерных взаимодействиях, рентген-эмульсионная камера
\newline \newline

The article considers events recorded in an experiment with X-ray emulsion chambers (XREC) to study the trunks of extensive air showers (EAS) with energies of nuclei of primary cosmic radiation (PCR) $E_0$~>~100~TeV, which currently have no explanation within the standard model of nuclear interactions (SM). The XREC method is the only one that allows studying EAS trunks with a high coordinate resolution of about 100 microns. EAS trunks contain the most complete information about primary acts of nuclear interactions, as a result of which all unusual events recorded by the XREC method are of great scientific interest. Several types of unusual events recorded by the XREC method were explained within the SM.\newline
}


\begin{document}

\maketitle
{\bf 1.~Введение}\newline
Исследования ПКИ высоких и сверхвысоких энергий с $E_0$ более сотни ТэВ позволяет изучать многие астрофизические аспекты, такие как природа источников космических лучей (КЛ), механизмы ускорения КЛ, возможные нарушения СМ и другие. Например, в ряде экспериментов делаются попытки обнаружить частицы странной кварковой материи в составе ПКИ~\cite{1}, или т.~н. странглеты, возможно имеющие отношение к темной материи.

Методы изучения ПКИ с $E_0$~>~100~ТэВ относятся к непрямым методам изучения КЛ и образованных ПКИ ШАЛ. Потоки ПКИ при данных $E_0$ относительно небольшие, и их характеристики изучаются моделированием ШАЛ, - электрон-фотонной и ядерно-активной компонент (ЭФК и ЯАК). Кроме того, характеристики ШАЛ, такие как количество электронов $N_\textrm{e}$ и мюонов $N_\mu$, сильно флуктуируют. Последнее затрудняет определение параметров первичного взаимодействия ядер ПКИ с атомами атмосферы. Однако в исследованиях~\cite{2,3} показано, что в стволе ШАЛ на расстоянии $R$~<~15~см от оси ливня флуктуации минимальны и регистрируемые характеристики ШАЛ максимально полно несут информацию об акте первичного взаимодействия ядер ПКИ с атмосферой.

Таким образом, цель экспериментов, позволяющих изучать структуру стволов ШАЛ, состоит не только в изучении, например, массового состава ПКИ~\cite{4}, но и в поиске необычных событий, возможно выходящих за рамки теории СМ. Примерами таких событий могут быть т.~н. выстроенные и проникающие события, зарегистрированные в РЭК. В частности, к таким событиям относятся событие JF~2af2, состоящее из $\sim$~40 $\gamma$-квантов, выстроенных вдоль прямой линии~\cite{5}, а также $\gamma$-адронное семейство СТРАНА, образованное ядром ПКИ с $E_0$~>~10~ПэВ и зарегистрированное в стратосферном эксперименте в Физическом институте им.~П.~Н.~Лебедева Российской академии наук под руководством К.~А.~Котельникова~\cite{6}. Для описания подобного рода событий, имеющих $p_{\perp}\sim$~10~ГэВ/с, необходима специальная теория.

Однако для разработки указанной выше теории необходима регистрация необычных событий, обоснованно указывающая на принадлежность подобных событий к данному типу частицы или взаимодействия, например, выстроенных, проникающих или образованных странглетами событий. В частности, до недавнего времени к необычным событиям, регистрируемым методом РЭК, относились гало, - большие пятна потемнений с площадями до 1000 и выше кв.~мм на рентгенографической пленке (РГП). Выполненными исследованиями было показано~\cite{3}, что данные события не относятся к необычным и образованы перекрытием функций пространственного распределения (ФПР) подпороговых относительно метода РЭК $\gamma$-квантов.

В настоящее время при изучении стволов ШАЛ в части поиска экзотики необходимы как регистрация необычных событий, так и доказательство, что данное событие невозможно объяснить в рамках СМ.

\begin{figure}[h]
\centering
\includegraphics[width=\linewidth]{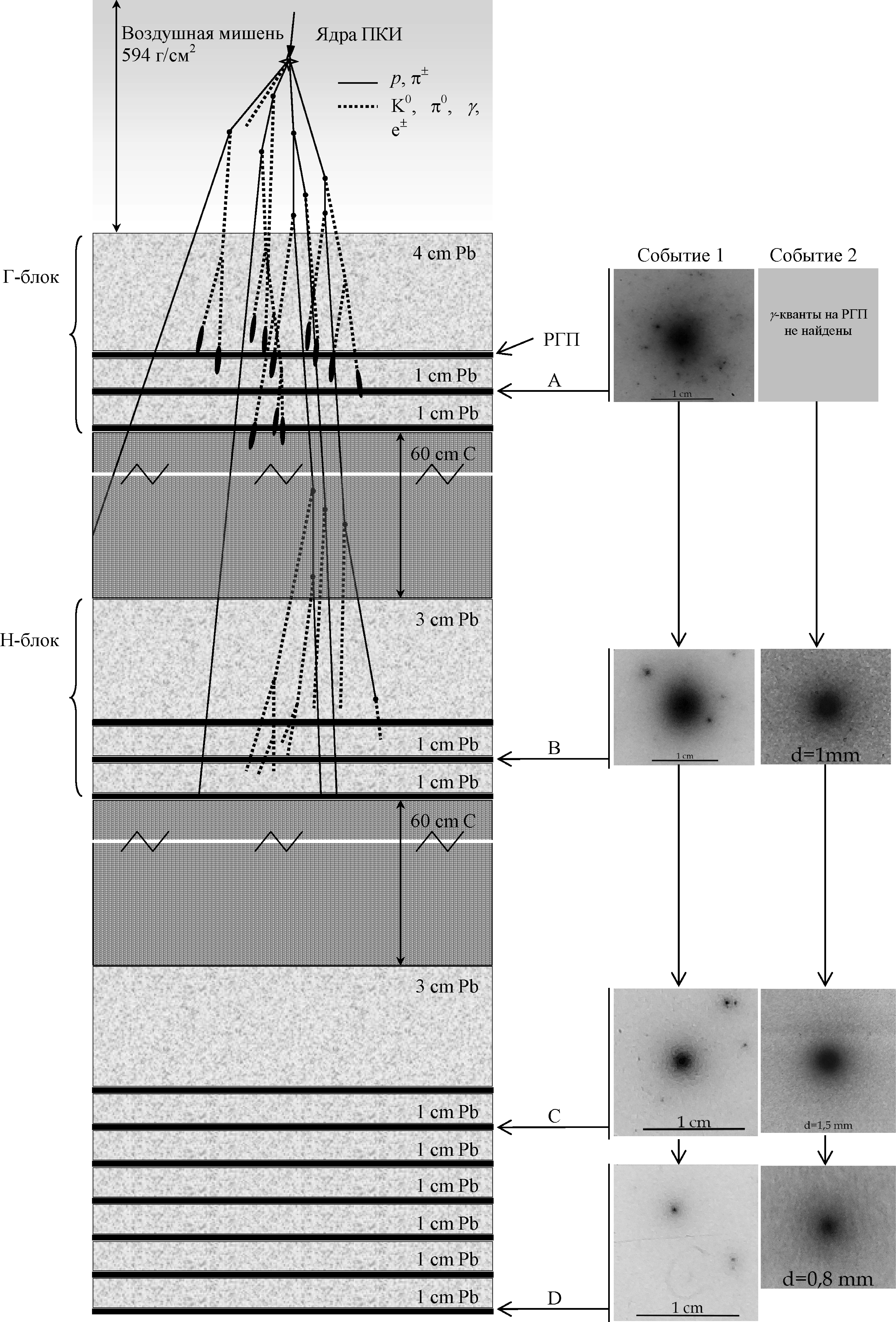}
\caption{Схема РЭК, а также примеры зарегистрированных на РГП т.~н. проникающих событий (события~1 и 2). Ряды РГП нумерованы сверху вниз от 1 до 13 с общей глубиной Pb 20~см.}
\label{fig1}
\end{figure}

{\bf 2.~Метод РЭК}\newline
Основная цель метода РЭК, - исследование стволов ШАЛ путем регистрации событий с высоким координатным разрешением. Одна из конфигураций экспериментальной установки РЭК, экспонировавшейся в эксперименте \mbox{ПАМИР} (594~г/см$^2$ стандартной атмосферы), приведена на рисунке~\ref{fig1}. РЭК состоит из Г-блока, где происходит основное поглощение и регистрация ЭФК, и адронного Н-блока для регистрации ЯАК ШАЛ, отделенного от Г-блока слоем углерода ($t_0^\textrm{Pb}$~=~6,4~г/см$^2$, $t_0^\textrm{C}$~=~43,3~г/см$^2$).

Как показано на рисунке~\ref{fig1}, первичные $\gamma$-кванты ЭФК ШАЛ, попадая в свинцовый поглотитель Pb, образуют вторичные каскады $\gamma'$-квантов, распределение которых по глубине Pb определяет площадь $S$ и степень потемнения $D$ пятен на РГП. РГП представляет собой подложку толщиной $\cong$~200~мкм, залитую с двух сторон фото-эмульсией толщиной $\cong$~30~x~2~мкм. $\gamma'$-кванты, проходя через микрокристаллы галоидного серебра AgHal, формируют латентное изображение, проявляемое и фиксируемое в процессе дальнейшей обработки РГП. В процессе экспозиции кристаллы AgHal абсорбируют $\gamma'$-кванты, энергия которых идет на образование металлического Ag. При наличии нескольких атомов металлического Ag в кристалле AgHal, последний фиксирует латентное изображение уже во всем своем объеме с коэффициентом усиления >~10$^7$ и вне зависимости от энергий $\gamma$-квантов. Таким образом, степень потемнения $D$ на РГП зависит только от потока вторичных $\gamma'$-квантов $n$ и миделевого сечения $s$ кристалла AgHal, характеризующего РГП, используемую в данном эксперименте:

\begin{equation}
D=4(1-e^{-ns})
\label{eq1}
\end{equation}

где $s$~=~(3,3~$\pm$~0,1)~мкм$^2$ для РГП в эксперименте \mbox{ПАМИР}, коэффициент~4 соответствует $D$($n\to\propto$). Переход от $D$ к энергии первичного $\gamma$-кванта $E_\gamma$ выполняется по калибровочным кривым от радиоактивных меток с известными характеристиками.

Экспонированная РГП просматривается с целью поиска и дальнейшей обработки отдельных пятен потемнения («отдельных $\gamma$-квантов») или групп пятен потемнения («семейств $\gamma$-квантов»). Минимальные значения потемнения и площади, $D_\textrm{th}$ и $S_\textrm{th}$, визуально наблюдаемые на РГП от отдельных $\gamma$-квантов имеют значения 0,05 и 0,1~мм$^2$ соответственно, что является визуальными порогами регистрации $\gamma$-квантов на РГП. Меньшие потоки $\gamma'$-квантов как отдельные $\gamma$-кванты на РГП визуально не наблюдаются, но со временем образуют вуаль и в дальнейшем фон потемнения $D_\textrm{ф}$, который учитывается при экспонировании и обработке РГП.

Плотности потемнений $D'$ от $\gamma$-квантов, измеренные фотометром на РГП с общим фоновым потемнением $D_\textrm{ф}$, пересчитаны в плотности потемнения $D_0'$ с учетом фона, как следует из формулы~\ref{eq1}:

\begin{equation}
D_0'=4(D'-D_\textrm{ф})/(4-D_\textrm{ф})
\label{eq2}
\end{equation}

и дальнейшие расчеты выполнены для значений плотностей потемнения $D\equiv D_0'$, пересчитанных с учетом фона РГП.

В случае, когда в данную точку на РГП приходит $N$-е количество $\gamma$-квантов с разными энергиями $E_{i~\gamma}$, дающими разные значения $D_i$, из формулы~\ref{eq1} следует суммарное потемнение:

\begin{equation}
D_\Sigma=4(1-\prod_{i=1}^{N}(1-D_i/4))
\label{eq3}
\end{equation}

или в случае равенства энергий $\gamma$-квантов $D_i$~=~$D$:

\begin{equation}
D_\Sigma^{(N)}=4(1-(1-D/4)^N)
\label{eq4}
\end{equation}

{\bf 3.~Моделирование эксперимента с РЭК}\newline
Моделирование экспериментальных данных, получаемых методом РЭК, требует как моделирования развития ШАЛ в атмосфере, так и прохождения ствола ШАЛ через РЭК. Характеристики ШАЛ на уровне наблюдения, полученные в модельном расчете, не должны противоречить данным РЭК, а также характеристикам ядерных взаимодействий, получаемых в экспериментах LHCf. Сравнительный анализ моделей ШАЛ и вопросы выбора модели ШАЛ для описания экспериментальных данных РЭК рассмотрены в~\cite{3}. Показано, что модель ШАЛ должна обеспечивать проведение расчетов с порогом по энергиям отслеживаемых частиц 100~ГэВ и ниже, а также порогом по $E_0$ до 3~ЭэВ и выше, описывать средний радиус семейств $\gamma$-квантов и соответствовать данным LHCf в диапазоне $x_\textrm{F}$~>~0,06, при том, что данные LHCf здесь немногочисленны и представлены узкими интервалами $\eta$~\cite{7}.

Модель ШАЛ, соответствующая перечисленным требованиям, разработана Р.~А.~Мухамедшиным~\cite{8} и получила название \mbox{FANSY}. В отличие от широко применяемых моделей ШАЛ комплекса \mbox{CORSIKA}~\cite{9}, в \mbox{FANSY} моделируются генерации более 300 адронов, что необходимо при изучении параметров стволов ШАЛ, регистрируемых методом РЭК.

Моделирование прохождения ЭФК ШАЛ через Pb РЭК требует значительных вычислительных мощностей. ФПР для вторичных e$^\pm$ и $\gamma'$-квантов, образованных первичными e$^\pm$ и $\gamma$-квантами различных энергий, получены моделированием в работах В.~В.~Учайкина~\cite{10} со значениями фитирования, приведенными в таблице~\ref{tabl:1}. Моделирование ФПР выполнено с учетом эффекта Ландау-Померанчука-Мигдалла (ЛПМ) и в осевом приближении (малые $r$ и большие энергии лавинных e$^\pm\gg \epsilon$(Pb)~=~7.4~МэВ, $N_\gamma$~=f($Er$)) на расстояниях $\sim$ длины когерентности. Эффект ЛПМ проявляется при энергиях $\gamma$-квантов >~10~ТэВ. Учет эффекта ЛПМ показывает рост пробега e$^\pm$ с энергией $\gamma$-квантов, что меняет формы каскадных кривых.

\setcounter{table}{0}
\renewcommand\thetable{\arabic{table}}
\begin{table*}[htbp]
\caption{Параметры ФПР е$^\pm$ и $\gamma$-квантов, полученных для РЭК в работах~\cite{10}.}
  \centering
    \begin{tabular}[t]{|l|c|c|}
\hline    
\multicolumn{1}{|c|}{Параметры ФПР}&min&max\\
\hline
Расстояние от оси каскада&1,58~мкм&6,56~см\\
\hline
Глубина свинцового поглотителя,~cu&1&60\\
\hline
Энергии первичных е$^\pm$ и $\gamma$-квантов&1~ГэВ&1~ПэВ\\
\hline
Потоки вторичных е$^\pm$ и $\gamma'$-квантов&0&$\sim$10$^{10}$~см$^{-2}$\\
\hline
    \end{tabular}
\label{tabl:1}    
\end{table*}

{\bf 4.~Необычные события, зарегистрированные в РЭК}\newline
На рисунке~\ref{fig1} приведены примеры двух групп необычных событий, зарегистрированных в РЭК. Семейства $\gamma$-квантов, относящиеся к первой группе событий, объяснение которым ранее не предлагалось, обозначены на рисунке~\ref{fig1} как «Событие~1». Ко второй группе событий, обозначенной на рисунке~\ref{fig1} как «Событие~2», отнесены события, объясняемые в исследованиях как события, образованные странной кварковой материей~\cite{1}. Оба события прослежены в рядах РГП на глубину 20~см Pb РЭК.

\begin{figure}[h]
\centering
\includegraphics[width=\linewidth]{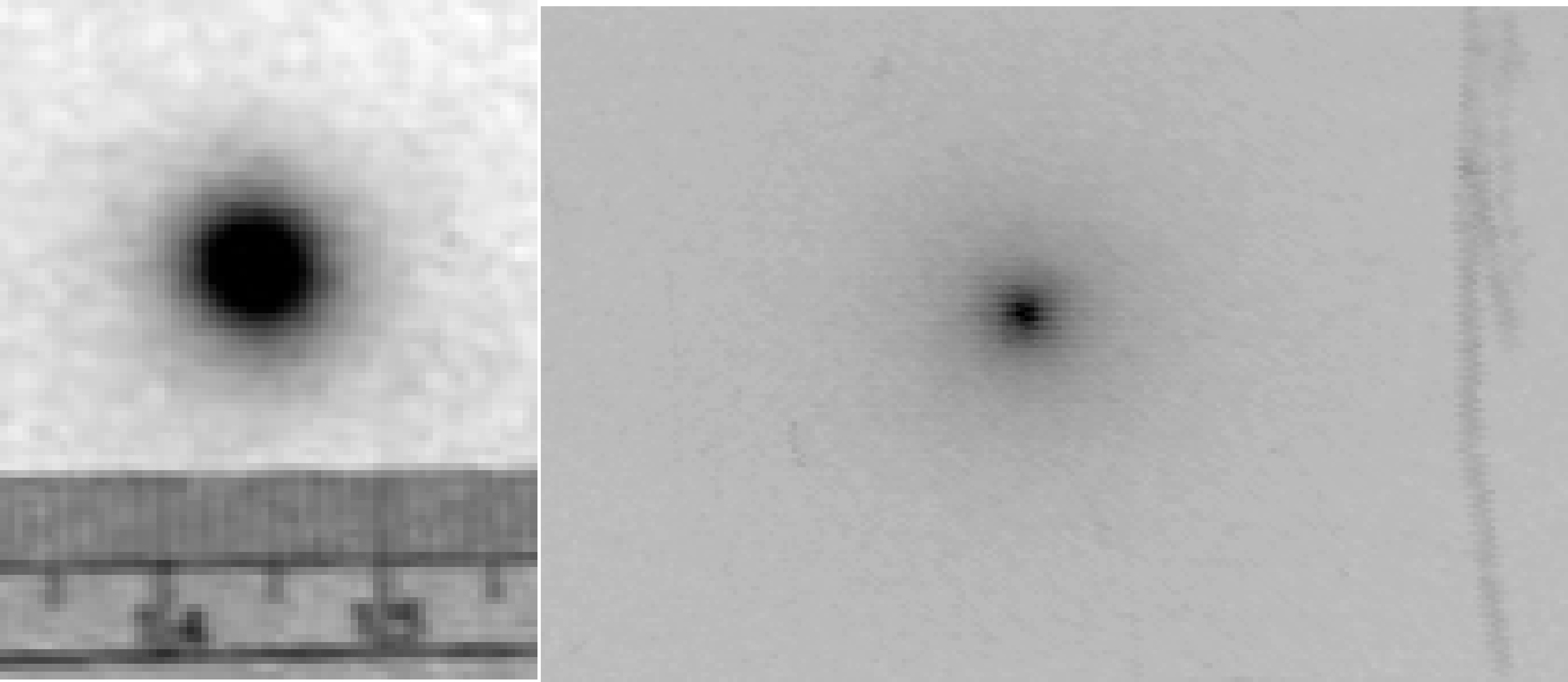}
\caption{Пятна потемнений на РГП от одиночных $\gamma$-квантов, зарегистрированные в стратосферном эксперименте \mbox{JACEE} (The Japanese-American Collaborative Emulsion Experiment) (5~г/см$^2$)~\cite{11} (слева) и на Тянь-Шаньской высокогорной научной станции (ТШВНС, эксперимент «Адрон», 695~г/см$^2$)~\cite{12} (справа).}
\label{fig2}
\end{figure}

Основные особенности события~2 состоят в том, что данное событие представляет собой пятно потемнения небольшой площади, а также в отсутствии $\gamma$-квантов, составляющих семейства $\gamma$-квантов. События, аналогичные событию~2, регистрировались также в экспериментах с РЭК на аэростатах и ТШВНС, примеры которых приведены на рисунке~\ref{fig2}.

Характеристики двух групп необычных событий, зарегистрированных в РЭК, приведены в таблице~\ref{tabl:2}: усредненное по площади диафрагмы фотометра потемнение $D$, потоки е$^\pm$ и $\gamma'$-квантов и площадь пятна потемнения $S$. Анализ данных, представленных в таблице~\ref{tabl:2}, показывает, что потоки $\gamma'$-квантов существенно меняются с глубиной Pb, - событие~1 затухает, электромагнитный каскад (ЭМК) в событии~2 развивается, достигает максимума на глубине $\sim$~30~cu и также затухает. Однако согласно теории странглетов~\cite{13}, последние должны генерировать ЭМК по всей глубине РЭК.

\renewcommand\thetable{\arabic{table}}
\begin{table*}[htbp]
\caption{Характеристики событий, зарегистрированных в эксперименте с РЭК, схема которой приведена на рисунке~\ref{fig1}.}
  \centering
    \begin{tabular}[t]{|c|c|c|c|c|c|c|c|c|}
\hline    
\multirow{2}{*}{№ ряда РГП}&\multicolumn{3}{c|}{"Событие~1"}&\multicolumn{3}{c|}{"Событие~2"}&\multirow{2}{*}{$h$,~г/см$^2$}&\multirow{2}{*}{$T$,~cu}\\
\cline{2-7}
&$D$&$n$,~мкм$^{-2}$&$S$,~мм$^2$&$D$&$n$,~мкм$^{-2}$&$S$,~мм$^2$&&\\
\hline                           
2&1,50&0,1444~$\pm$~0,0008&66&-&-&-&57&8,9\\
\hline
4&-&-&-&0,9~$\pm$~0,1&0,07~$\pm$~0,00&0,8&162&17,3\\
\hline
5&1,77&0,1792~$\pm$~0,0009&49&1,3~$\pm$~0,2&0,12~$\pm$~0,02&0,8&173&19,1\\
\hline
6&-&-&-&1,7~$\pm$~0,2&0,17~$\pm$~0,03&1,1&185&20,9\\
\hline
7&-&-&-&1,8~$\pm$~0,2&0,18~$\pm$~0,03&1,3&279&27,6\\
\hline
8&1,61~$\pm$~0,07&0,158~$\pm$~0,009&10&1,9~$\pm$~0,2&0,20~$\pm$~0,03&1,8&290&29,4\\
\hline
9&-&-&-&2,2~$\pm$~0,2&0,24~$\pm$~0,03&1,8&301&31,1\\
\hline
10&-&-&-&2,0~$\pm$~0,2&0,21~$\pm$~0,02&1,8&313&32,9\\
\hline
11&-&-&-&1,5~$\pm$~0,2&0,14~$\pm$~0,02&1,1&324&34,7\\
 \hline
12&-&-&-&1,3~$\pm$~0,2&0,12~$\pm$~0,02&0,8&335&36,4\\
 \hline
13&0,60~$\pm$~0,05&0,050~$\pm$~0,004&1&0,9~$\pm$~0,1&0,08~$\pm$~0,01&0,5&347&38,2\\
 \hline
    \end{tabular}
\label{tabl:2}    
\end{table*}

\begin{figure}[h]
\centering
\includegraphics[width=\linewidth]{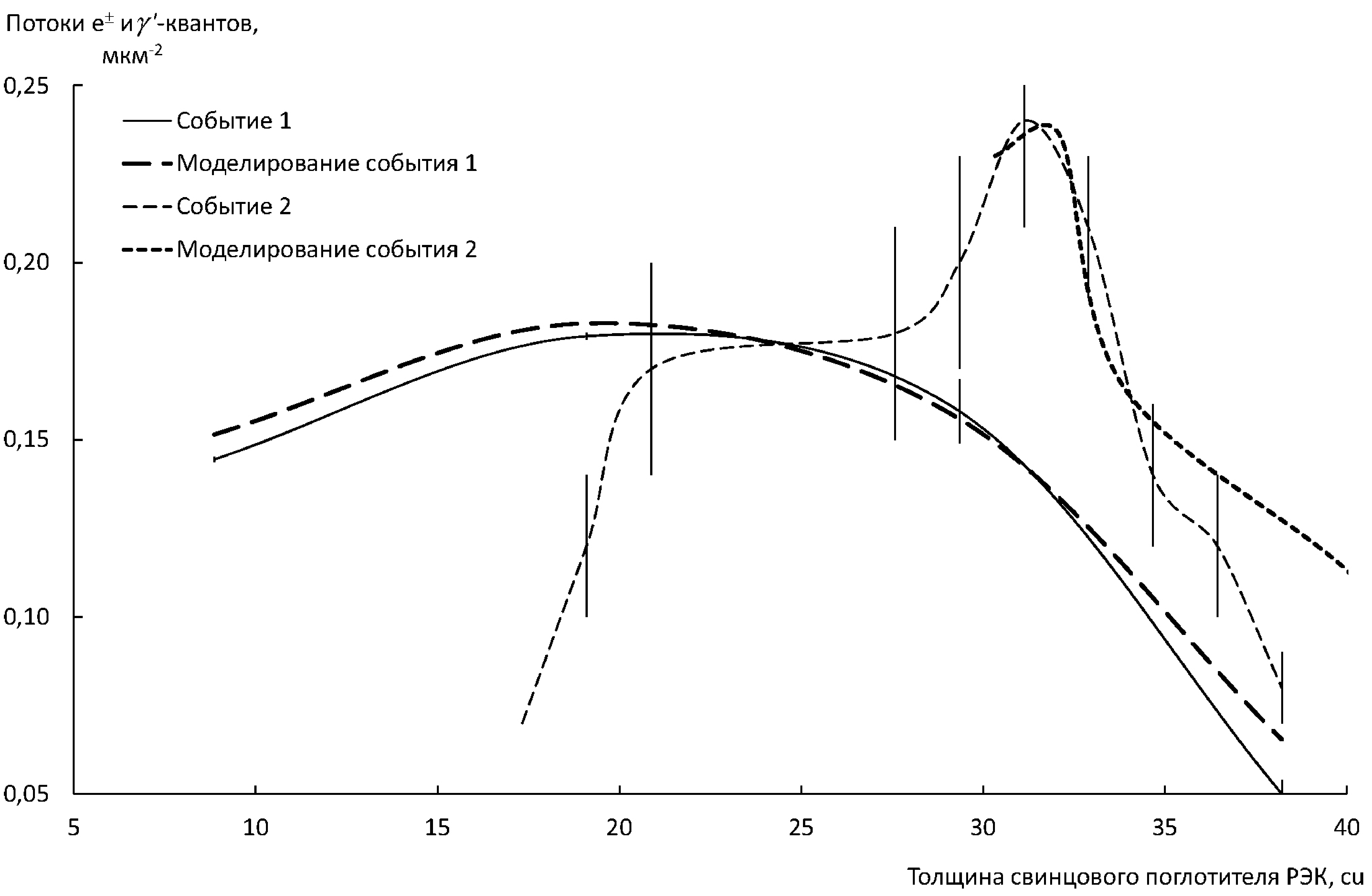}
\caption{Зависимости потоков вторичных е$^\pm$ и $\gamma'$-квантов экспериментальных и модельно полученных событий от толщины свинцового поглотителя РЭК.}
\label{fig3}
\end{figure}

{\bf 5.~Природа необычных событий, зарегистрированных в РЭК}\newline
Результаты моделирования значений потоков вторичных е$^\pm$ и $\gamma'$-квантов приведены на рисунке~\ref{fig3} для рассмотренных выше событий~1 и 2 (см. таблицу~\ref{tabl:2}). Моделирование выполнено в рамках СМ без привлечения специальных теорий.

Результаты моделирования события~1 показали практически идентичные кривые потемнения, полученные как экспериментально в РЭК, так и в модели. Расчеты показали, что событие~1 образовано $\sim$~30-40~тыс. подпороговыми $\gamma$-квантами, $\sim$~120 $\gamma$-квантами с энергиями до 10~ТэВ и $\sim$~30 $\gamma$-квантами с энергиями >~100~ТэВ. Как следует из каскадной теории, полученная зависимость $log(N_\gamma)$~=~f($log(E_\gamma)$) линейна с $R_a^2$~>~95~$\%$ и соответствует результатам моделирования распределения е$^\pm$ и $\gamma$-квантов в ШАЛ по энергиям, выполненного в~\cite{3}.

Основываясь на результатах моделирования события~1 и большой статистике модельных параметров ШАЛ, полученных в разработанной специально для экспериментов с РЭК модели \mbox{FANSY}~\cite{8}, можно с высокой степенью определенности утверждать, какой тип ядра ПКИ образовал ШАЛ, ствол которого был зарегистрирован в РЭК как событие~1. Для решения данной задачи необходимо обратить внимание, что энергии и потоки высокоэнергичных $\gamma$-квантов в стволе ШАЛ являются маркерами необычных событий, в частности позволяющими отнести ядра к легкой (\textit{p}+He) или тяжелой (>~He) группе ПКИ как в событиях, показанных на рисунке~\ref{fig1}.

В случае события~1, один-два $\gamma$-кванта в стволе ШАЛ должны иметь энергии $\sim$~300~ТэВ, - меньшие энергии дадут расчетные потоки вторичных е$^\pm$ и $\gamma'$-квантов ниже экспериментальных, более высокие энергии $\gamma$-квантов, - выше экспериментально полученных. В соответствии с результатами моделирования стволов ШАЛ, регистрируемых в РЭК, событие~1 образовано ядром, относящимся к легкой компоненте ПКИ \textit{p}+He, в данном случае протоном, так как согласно расчетам максимальные энергии $\gamma$-квантов, образованные ядрами тяжелой компоненты ПКИ, существенно меньше даже 200~ТэВ. Также из моделирования ШАЛ с применением \mbox{FANSY} получено, что энергия ядра легкой компоненты ПКИ, образовавшего событие~1, составляет $\sim$~60~ПэВ.

Природа события~2 более сложная, нежели события~1. Причина состоит в том, что пятна потемнения на РГП в событии~2 имеют заметно меньшую площадь, а значит образованы $\gamma$-квантами с меньшими энергиями. Однако ЭМК от данных $\gamma$-квантов также проходят на большую глубину Pb, то есть энергии $\gamma$-квантов должны быть большими. Но одновременно вторичные ЭМК $\gamma'$-квантов в первых рядах РГП не зарегистрированы. Перечисленные выше противоречивые условия могут реализоваться, когда ЭМК ствола ШАЛ состоит из относительно небольшого количества низко-энергичных $\gamma$-квантов, которые полностью поглотились в Г-блоке РЭК. При этом в Н-блоке РЭК образовались один-два высоко-энергичных $\gamma$-кванта, которые образуют ЭМК, идущий вглубь РЭК. Очевидно, что данный каскад будет образовывать пятна потемнений малых площадей и без «семейств $\gamma$-квантов».

Моделирование показало, что начало вторичного ЭМК события~2 должно быть образовано ЯАК на глубине $\sim$~13~cu РЭК, или в первом слое углерода толщиной 60~см. В частности, в~\cite{14} показано, что пробег неупругого взаимодействия протонов $\lambda_\textrm{inel}$~=~115–120~г/см$^2$, что соответствует началу Н-блока на глубине 128~г/см$^2$ РЭК. В силу сказанного, с высокой вероятностью данный ШАЛ, ствол которого зарегистрирован в РЭК как событие~2, образован вторичным протоном.

\begin{figure}[h]
\centering
\includegraphics[width=\linewidth]{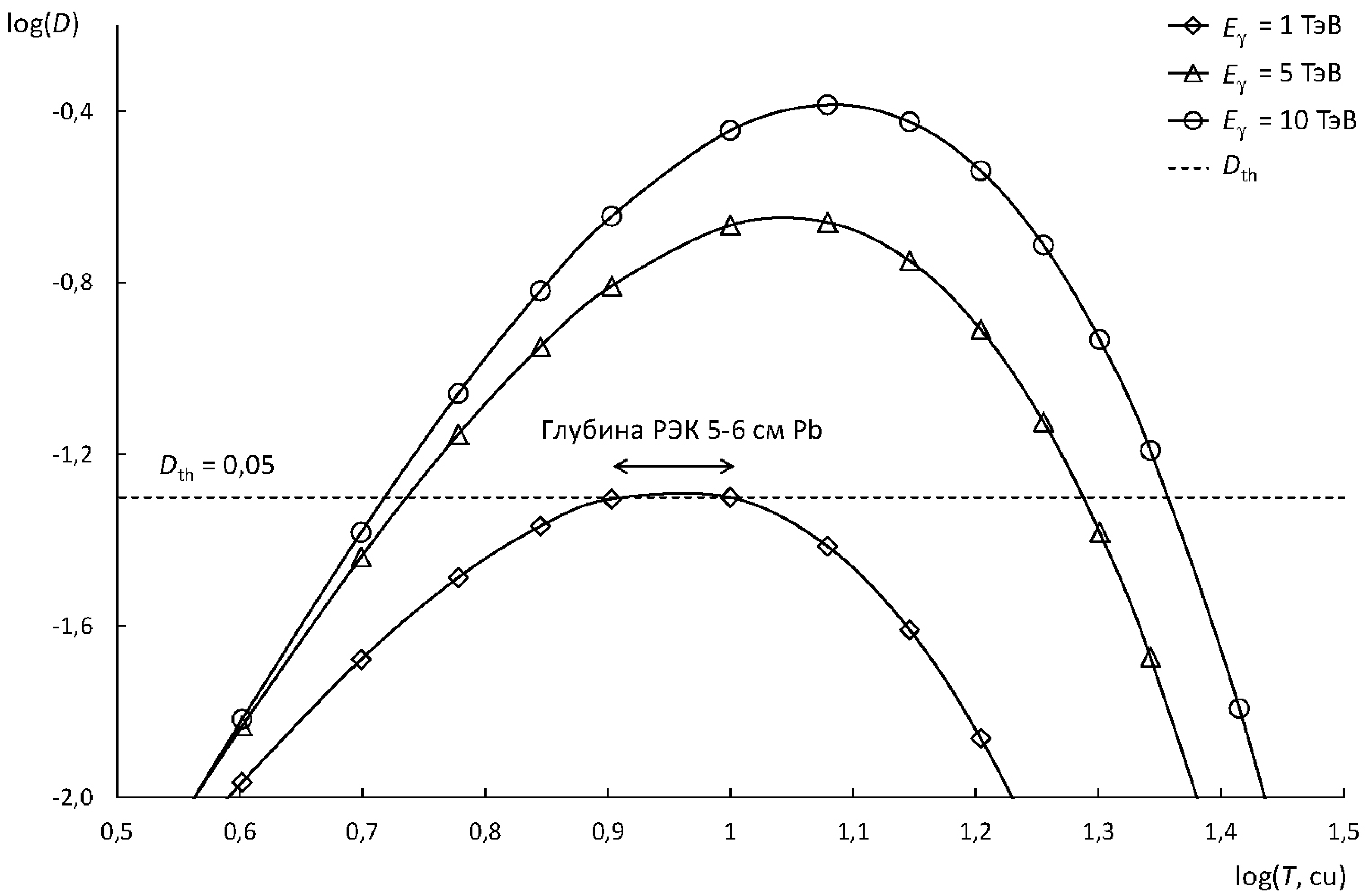}
\caption{Потемнение на РГП, вызванное ЭМК от первичного $\gamma$-кванта, в зависимости от глубины РЭК. Порог видимости $\gamma'$-кванта показан горизонтальным пунктиром, соответствующим $D_{th}$~=~0,05. Расстояние от первичного $\gamma$-кванта $R$~=~208~мкм (значение, фитированное в модельном расчете~\cite{10}).}
\label{fig4}
\end{figure}

Рассмотрим каскадные кривые вторичных $\gamma'$-квантов различных энергий в РЭК. На рисунке~\ref{fig4} показано потемнение на РГП, вызванное ЭМК от первичного $\gamma$-кванта, в зависимости от глубины Pb. Чтобы $\gamma'$-квант был виден при визуальной обработке РГП, площадь пятна потемнения $S$ должна быть >~0,1~мм$^2$ с плотностью потемнения $D$~>~0,05. Уровень видимость $\gamma'$-квантов на РГП, соответствующий $D$~=~0,05 формируется потоком е$^\pm$ и $\gamma'$-квантов $N_{e^\pm,\gamma}$~=~0,004~мкм$^{-2}$ на расстоянии $R$~=~178~мкм от оси первичного $\gamma$-кванта (см. формулу~\ref{eq1}).

Анализ рисунка~\ref{fig4} показывает, что минимальная энергия одиночного $\gamma$-кванта, который может быть визуально зарегистрирован методом РЭК, составляет $\sim$~1~ТэВ. С ростом $E_\gamma$ каскадные кривые сдвигаются вглубь РЭК, а уровень потемнения $D$ увеличивается. Из рисунка~\ref{fig4} также следует, что на определенной глубине Pb РЭК будет наблюдаться избыток визуально регистрируемых $\gamma$-квантов. В случае 1~ТэВ-го $\gamma$-кванта, - это 5-6~см Pb, или 2-3 ряды РГП. Очевидно, что глубина, на которой будет наблюдаться указанный избыток $\gamma$-квантов, определяется распределением $\gamma$-квантов по $E_\gamma$ в регистрируемых стволах ШАЛ и массовым составом ПКИ при данном пороге по энергии экспериментальной установки, который в методе РЭК связан со средним радиусом семейств $\gamma$-квантов, равным (1,96~$\pm$~0,06)~см. В частности, при $E_0$~=~5~ПэВ средневзвешенные средние радиусы семейств $\gamma$-квантов стволов ШАЛ, образованных протонами и ядрами Fe, различаются в два раза~\cite{3}.

{\bf 6.~Обсуждение}\newline
В экспериментах с РЭК регистрируются разнообразные события, не все из которых сразу находят свое объяснение. Например, пятно потемнения на РГП большой площади (гало) 745~мм$^2$ по изоденсе $D$~=~0,5 c $\Sigma E_\gamma \sim$~6~ПэВ, названное «Андромеда», зарегистрированное в 1970~г. на установке в горах Чакалтая (520~г/см$^2$), природа которого долгое время оставалась необъясненной. Только много позднее был показан механизм образования гало таких и существенно больших площадей~\cite{3} без привлечения т.~н. «экзотики», но в рамках СМ.

С другой стороны и в силу сказанного нестандартные модели, в частности модель странной кварковой материи, весьма удобны, так как подобными моделями можно объяснить практически все наблюдаемые в РЭК события. В модели прохождения странглета через РЭК, последний, проходя на любую глубину Pb, излучает $\Lambda$-гипероны, продукты распада которых регистрируются на РГП как $\gamma'$-кванты. Это объясняет, например, регистрацию проникающих событий в глубокой свинцовой камере. Однако существование странглетов необходимо подтвердить экспериментально. В частности, поиск странглетов определен как одна из научных целей в исследованиях КЛ на ТШВНС~\cite{12} (38~Всероссийская конференция по космическим лучам ВККЛ-2024, ФИАН).

Очень вероятно, что существуют простые объяснения природы наблюдаемых в экспериментах с РЭК необычных событий, не выходящие за рамки СМ. Одно из таких объяснений рассмотрено в данной статье и связано с особенностями развития и регистрации ЭМК в Pb РЭК. Очевидное преимущество данной точки зрения состоит в том, что для описания событий, считающихся необычными, не требуются новые теории. Необходимы только достаточные вычислительные мощности, которых несколько десятков лет назад не было. Например, для расчета гало необходим учет очень больших потоков подпороговых относительно метода РЭК $\gamma'$-квантов~\cite{3}, а фитирование ФПР $\gamma'$-квантов для экспериментов с РЭК~\cite{10} представляло собой весьма трудоемкую задачу, выполнение которой было возможно только на больших вычислительных комплексах.\newline

{\bf 7.~Выводы}\keepwithnext
\begin{enumerate}
\item Объяснена природа двух типов необычных событий в стволах ШАЛ, зарегистрированных методом РЭК и прошедших на большую глубину Pb РЭК: событие с семействами $\gamma$-квантов и событие, представляющее собой равномерное пятно потемнения на РГП без отдельных $\gamma'$-квантов. Показано, что первое событие образовано несколькими высоко-энергичными $\gamma$-квантами, второе событие образовано вторичным протоном, образовавшем ЭМК в глубине РЭК.

\item Результаты экспериментов с РЭК достаточно сложны для описания в методической части. В экспериментах с РЭК можно получить много необычных событий, включая и, возможно, выходящие за рамки СМ. Однако для целей дальнейших исследований необычные события должны быть тщательно обработаны и изучены.

\item Эксперименты с РЭК позволяют изучать стволы ШАЛ с высоким координатным разрешением, недостижимым в других экспериментах, и где информация о первичных ядерных взаимодействиях минимально искажена. Дальнейшая обработка экспериментальных данных РЭК перспективна, т.~к. позволяет получать новые результаты на стыке астрофизики и ядерно-физического аспекта ПКИ.
\end{enumerate}

\end{document}